\documentclass[preprint,twocolumn]{aastex61}

\usepackage{subfigure}
\usepackage{url}
\usepackage{comment}
\usepackage{color}
\usepackage{enumitem}

\setlist[enumerate]{noitemsep}
\setlist[enumerate,1]{leftmargin=*}
\setlist[itemize]{noitemsep}
\setlist[itemize,1]{leftmargin=*}
\setlist[description]{noitemsep}
\setlist[description,1]{leftmargin=*}
\setenumerate[0]{label=(\arabic*)}

\shorttitle{Neutron Capture Abundances of a New $r$-II Star}
\shortauthors{Sakari et al.}

\begin{document}

\title{The $r$-Process Pattern of a Bright, Highly $r$-Process-Enhanced Metal-Poor Halo Star at $[\rm{Fe/H}] \sim -2$}

\author{Charli M. Sakari}
\affil{Department of Astronomy, University of Washington, Seattle WA
98195-1580, USA}

\author{Vinicius M. Placco}
\affil{Department of Physics, University of Notre Dame, Notre Dame, IN 46556,  USA}
\affil{Joint Institute for Nuclear Astrophysics Center for the
  Evolution of the Elements (JINA-CEE), USA}

\author{Terese Hansen}
\affil{Observatories of the Carnegie Institution of Washington, 813
  Santa Barbara Street, Pasadena, CA 91101, USA}

\author{Erika M. Holmbeck}
\affil{Department of Physics, University of Notre Dame, Notre Dame, IN 46556,  USA}
\affil{Joint Institute for Nuclear Astrophysics Center for the
  Evolution of the Elements (JINA-CEE), USA}

\author{Timothy C. Beers}
\affil{Department of Physics, University of Notre Dame, Notre Dame, IN 46556,  USA}
\affil{Joint Institute for Nuclear Astrophysics Center for the Evolution of the Elements
(JINA-CEE), USA}

\author{Anna Frebel}
\affil{Department of Physics and Kavli Institute for Astrophysics and
  Space Research, Massachusetts Institute of Technology, Cambridge, MA
  02139, USA}

\author{Ian U. Roederer}
\affil{Department of Astronomy, University of Michigan, 1085 S. University Ave., Ann Arbor, MI 48109, USA}
\affil{Joint Institute for Nuclear Astrophysics Center for the Evolution of the Elements
(JINA-CEE), USA}

\author{Kim A. Venn}
\affil{Department of Physics and Astronomy, University of Victoria,
  Victoria, BC, Canada}

\author{George Wallerstein}
\affil{Department of Astronomy, University of Washington, Seattle WA
98195-1580, USA}

\author{Christopher Evan Davis}
\affil{Department of Astronomy, University of Washington, Seattle WA
98195-1580, USA}

\author{Elizabeth M. Farrell}
\affil{Department of Astronomy, University of Washington, Seattle WA
98195-1580, USA}

\author{David Yong}
\affil{Research School of Astronomy and Astrophysics, Australian National University, Canberra, ACT 2611, Australia}

\correspondingauthor{Charli M. Sakari}
\email{sakaricm@u.washington.edu}



\begin{abstract}
A high-resolution spectroscopic analysis is presented for a new
highly $r$-process-enhanced ($[\rm{Eu/Fe}]~=~1.27$,
$[\rm{Ba/Eu}]~=~-0.65$), very metal-poor ($[\rm{Fe/H}]~=~-2.09$),
retrograde halo star, RAVE J153830.9-180424, discovered as part of the
R-Process Alliance survey.  At $V=10.86$, this is the brightest and
most metal-rich $r$-II star known in the Milky Way halo.  Its
brightness enables high-S/N detections of a wide variety of chemical
species that are mostly created by the $r$-process, including some
infrequently detected lines from elements like Ru, Pd, Ag, Tm, Yb, Lu,
Hf, and Th, with upper limits on Pb and U.  This is the most complete
$r$-process census in a very metal-poor $r$-II star. J1538-1804 shows no
signs of $s$-process contamination, based on its low [Ba/Eu] and
[Pb/Fe].  As with many other $r$-process-enhanced stars, J1538-1804's
$r$-process pattern matches that of the Sun for elements between the
first, second, and third peaks, and does not exhibit an actinide
boost.  Cosmo-chronometric age-dating reveals the $r$-process material
to be quite old.  This robust main $r$-process pattern is a necessary
constraint for $r$-process formation scenarios (of particular interest
in light of the recent neutron star merger, GW~170817), and has
important consequences for the origins of $r$-II stars.  Additional
$r$-I and $r$-II stars will be reported by the R-Process Alliance in
the near future.
\end{abstract}


\section{Introduction}\label{sec:Intro}

The very metal-poor stars ($[\rm{Fe/H}] < -2$) are believed to be some
of the oldest objects in the Milky Way (MW).  These stars retain the
chemical signatures of the few stars that evolved and died before them
(e.g., \citealt{Frebel2015}).  The subset of stars that are highly
enhanced in the heavy elements that form via the rapid ($r$-) neutron
capture process are of particular interest, as their abundances trace
the yields from early $r$-process events.  The signatures of the
$r$-process are seen throughout the Galaxy \citep{Roederer2013}, but
$r$-process-enhanced stars enable measurements of a wide assortment of
neutron-capture elements.  These stars are classified according to
their Eu abundances: $r$-I stars have $+0.3<[\rm{Eu/Fe}]\le+1$, while
$r$-II stars have $[\rm{Eu/Fe}]>+1$
\citep{Christlieb2004}.\footnote{Both have $[\rm{Ba/Eu}]<0$ to
  minimize contamination from the slow ($s$-) process.} There are
presently only $\sim30$ $r$-II and $\sim100$ $r$-I stars known (see
the JINAbase compilation; \citealt{Abohalima2017}). Studies of these
stars have found a nearly identical main $r$-process pattern (for Ba
to Hf) in all types of stars and in all environments, with variations
among the lightest and heaviest elements (e.g.,
\citealt{Sneden1994,Roederer2014b,SiqueiraMello2014,Ji2016,Placco2017}).

Neutron star mergers (NSMs) have long been suspected to be a site of
the $r$-process (e.g.,
\citealt{LattimerSchramm1974,Rosswog2014,Lippuner2017}).  The recent
detection of GW~170817 \citep{Abbott2017} and subsequent $r$-process
nucleosynthesis (e.g., \citealt{Chornock2017}) demonstrate
that these conditions can indeed be met in NSMs.  Galactic chemical
evolution models (e.g., \citealt{Cote2017}) also suggest
that NSMs can produce all the observed Eu in the MW.  However,
problems such as coalescence time and NSM rates still remain,
prompting the question: Is there only a single site for $r$-process
nucleosynthesis, and are the physical conditions always identical?
Standard core-collapse supernovae seem to have been ruled out as the
site for most of the $r$-process elements (though they may form the
lighter elements; e.g., \citealt{ArconesThielemann2013}), but the
magneto-rotational supernovae (e.g., \citealt{Winteler2012}) remain
another option. Observations of the detailed $r$-process pattern in
large samples of stars will be useful for constraining the fundamental
physics and site(s) behind the $r$-process, through determinations of
the relative abundances of second- vs. third-peak elements, the
presence of actinide boosts (e.g., \citealt{Schatz2002}), and the
behavior of the first-peak elements.

Also important is the rate at which $r$-process events occur, as well as
where and when.  The detection of $r$-process-enhanced stars in
ultra-faint dwarf galaxies \citep{Ji2016}, the age-dating of
$r$-process-enhanced stars \citep{Sneden1996,Cayrel2001}, trends with
metallicity or other chemical abundances
\citep{MaciasRamirezRuiz2016}, the relative numbers of $r$-I and $r$-II
stars \citep{Barklem2005}, and the amount of $r$-process material in a
given environment (e.g.,
\citealt{TsujimotoNishimura2015,Beniamini2016}) are all important for
deciphering the site of the $r$-process.  Obtaining high-precision,
detailed abundance patterns and understanding the $r$-process-enhanced
stars as a stellar population in a statistical sense requires a much
larger sample of $r$-I and $r$-II stars.

The R-Process Alliance is a collaboration with the aim of identifying
the site(s) of the $r$-process.  The first phase of the Alliance is
dedicated to discovering larger samples of $r$-I and $r$-II stars in
the Milky Way through medium- and high-resolution spectroscopy (Placco
et al.,  Hansen et al., Sakari et al., {\it in prep.}).  This letter
presents the detailed $r$-process abundances of an $r$-II star,
RAVE~J153830.9-180424 (hereafter J1538-1804) that was discovered in
the northern hemisphere sample of Sakari et al. ({\it in prep.},
hereafter Paper~I).  Future papers will present additional $r$-I and
$r$-II stars discovered by the R-Process Alliance.

\vspace*{-0.1in}
\section{Observations, Data Reduction, and Analysis Techniques}\label{sec:Observations}
J1538-1804 was identified as a metal-poor star in the re-analyzed data
from the RAdial Velocity Experiment (RAVE) survey {\color{red}
  \citep{RAVEref}} by \citet{Matijevic2017}.  The star was followed up
at medium-resolution ($R\sim 2000$) in the blue ($3300-5000$ \AA) to
determine atmospheric parameters (Placco et al., {\it in prep.}), and
was subsequently identified as an $r$-II star in the northern
hemisphere, high-resolution component of the R-Process Alliance
(Paper~I), based on echelle spectroscopy with the Astrophysical
Research Consortium (ARC) 3.5~-~m telescope at Apache Point
Observatory (APO). The target was then followed up at higher
resolution ($R\sim83,000$ in the blue and $R\sim65,000$ in the red),
with the $0.\arcsec35$ slit and $1\times 1$ binning) on 4 May, 2017
with the Magellan Inamori Kyocera Echelle (MIKE) spectrograph
\citep{Bernstein2003} on the Magellan-Clay Telescope at Las Campanas
Observatory.  A wavelength coverage of $3200-5000$
\AA \hspace{0.025in} was obtained in the blue, and $4900-10000$
\AA \hspace{0.025in} in the red.  The details for the MIKE
observations are listed in Table \ref{table:Targets}; the conditions
were photometric, with 0.4\arcsec \hspace{0.025in} seeing. The spectra
were reduced using the Image Reduction and Analysis Facility program
(IRAF)\footnote{IRAF is distributed by   the National Optical
  Astronomy Observatory, which is operated by the Association of
  Universities for Research in Astronomy, Inc., under cooperative
  agreement with the National Science Foundation.} and the {\sc mtools}
package.\footnote{\url{http://www.lco.cl/telescopes-information/magellan/instrum
ents/mike/iraf-tools/iraf-mtools-package}}

All abundances were derived with spectrum syntheses, using the 2017
version of {\tt MOOG} \citep{Sneden} with an appropriate treatment of
scattering
\citep{Sobeck2011}.\footnote{\url{https://github.com/alexji/moog17scat}}
The atmospheric parameters of the stars were determined by flattening
trends in Fe I lines with wavelength, reduced equivalent width, and
excitation potential (EP), and by forcing agreement between Fe I and Fe II
abundances.  For each \ion{Fe}{1} line, a $<$3D$>$, non-Local
Thermodynamic Equilibrium (NLTE) correction \citep{Amarsi2016} was
applied to the LTE abundance.  Paper~I will demonstrate that the NLTE
parameters are generally in better agreement with photometric
temperatures; however, at $[\rm{Fe/H}]\sim -2$, the differences
between NLTE and LTE atmospheric parameters are generally negligible.
Using LTE atmospheric parameters also has a slight impact on the
derived abundances; see Table~\ref{table:Abunds}.

The line lists were generated with the {\tt linemake}
code\footnote{\vspace{0.5in}\url{https://github.com/vmplacco/linemake}}
(C. Sneden, {\it private comm.}) with additions from
\citet{Cowan2005}, \citet{Xu2007}, and \citet{Sneden2009}, and include
hyperfine structure, isotopic splitting, and molecular lines from CH,
C$_{2}$, and CN.  The atmospheric parameters and abundances of the
light and iron-peak elements are provided in Paper~I, though Table
\ref{table:Targets} lists the final, adopted atmospheric parameters,
the [C/Fe], and the average [$\alpha$/Fe].  Note that this star is not
C-enhanced.  This star was also included in the first Gaia data
release \citep{GAIAref}.  Its proper motion, parallax, and velocity
demonstrate that it is a retrograde halo star (see Table
\ref{table:Targets}), even accounting for the large uncertainty in the
parallax.

\clearpage

\begin{deluxetable}{@{}cc}
\tabletypesize{\scriptsize}
\tablecolumns{2}
\tablewidth{6in}
\tablecaption{Target Information\label{table:Targets}}
\hspace*{-2in}
\tablehead{
Parameter & Value}
\startdata
Aliases                 & TYC~6189-285-1\\
                        & 2MASS J15383085-1804242 \\
RA (J2000)              & 15:38:30.85 \\
Dec (J2000)             & -18:04:24.2 \\
$V$                     & 10.86       \\
$K$                     &  8.484      \\
Observation Date        & 4 May 2017 (MJD = 57877) \\
Exposure Time (s)  & 1800 \\
S/N, 3700 \AA$^{a}$ & 287 \\
$v_{\rm{helio}}$ (km s$^{-1}$) & $131.3\pm0.5$ \\
$U$ (km s$^{-1}$)$^{b}$  & $64^{+6874}_{-104}$ \\
$V$ (km s$^{-1}$)$^{b}$  & $-774^{+536}_{-4725}$ \\
$W$ (km s$^{-1}$)$^{b}$  & $147^{+85}_{-8776}$ \\
$T_{\rm{eff}}$ (K)        & $4752\pm30$ \\
$\log g$                & $1.63\pm0.1$ \\
$\xi$ (km/s)            & $1.51\pm0.15$ \\
$[$Fe/H$]$              & $-2.09\pm0.02$\\
$[$C/Fe$]^{c}$          & $+0.26\pm0.10$ \\
$[\alpha$/Fe$]^{d}$     & $+0.34\pm0.03$ \\
\enddata
\tablenotetext{a}{S/N is per resolution element}
\tablenotetext{b}{The large uncertainty is due to the uncertain Gaia DR1 parallax.}
\tablenotetext{c}{This is the ``natal'' [C/Fe], calculated with the evolutionary corrections of \citet{Placco2014c}.  The measured [C/Fe] is -0.10.}
\tablenotetext{d}{This is an average of [Mg/Fe], [Si/Fe], and [Ca/Fe]
  from Paper~I}
\end{deluxetable}

\section{$r$-Process Patterns}\label{sec:Abunds}
\begin{figure*}[h!]
\begin{center}
\includegraphics[scale=0.6,clip,trim=0.5in 0 0 0]{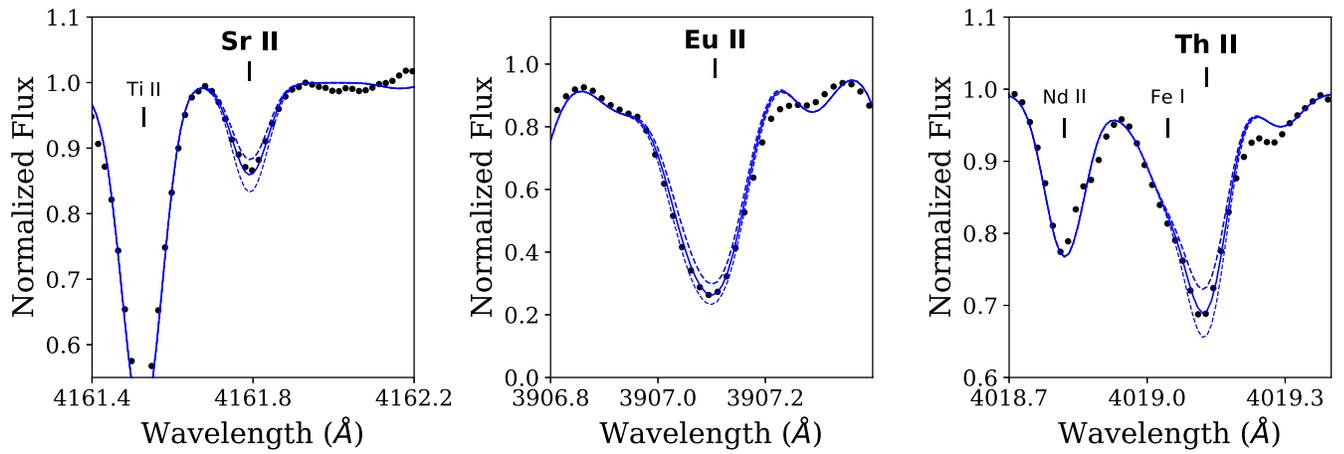}
\caption{Best-fit syntheses to Sr, Eu, and Th lines (solid lines), along with
  $\pm0.1$ dex abundances (dashed lines).\label{fig:Spectra}}
\end{center}
\end{figure*}

\begin{figure*}[h!]
\begin{center}
\centering
\includegraphics[scale=0.75]{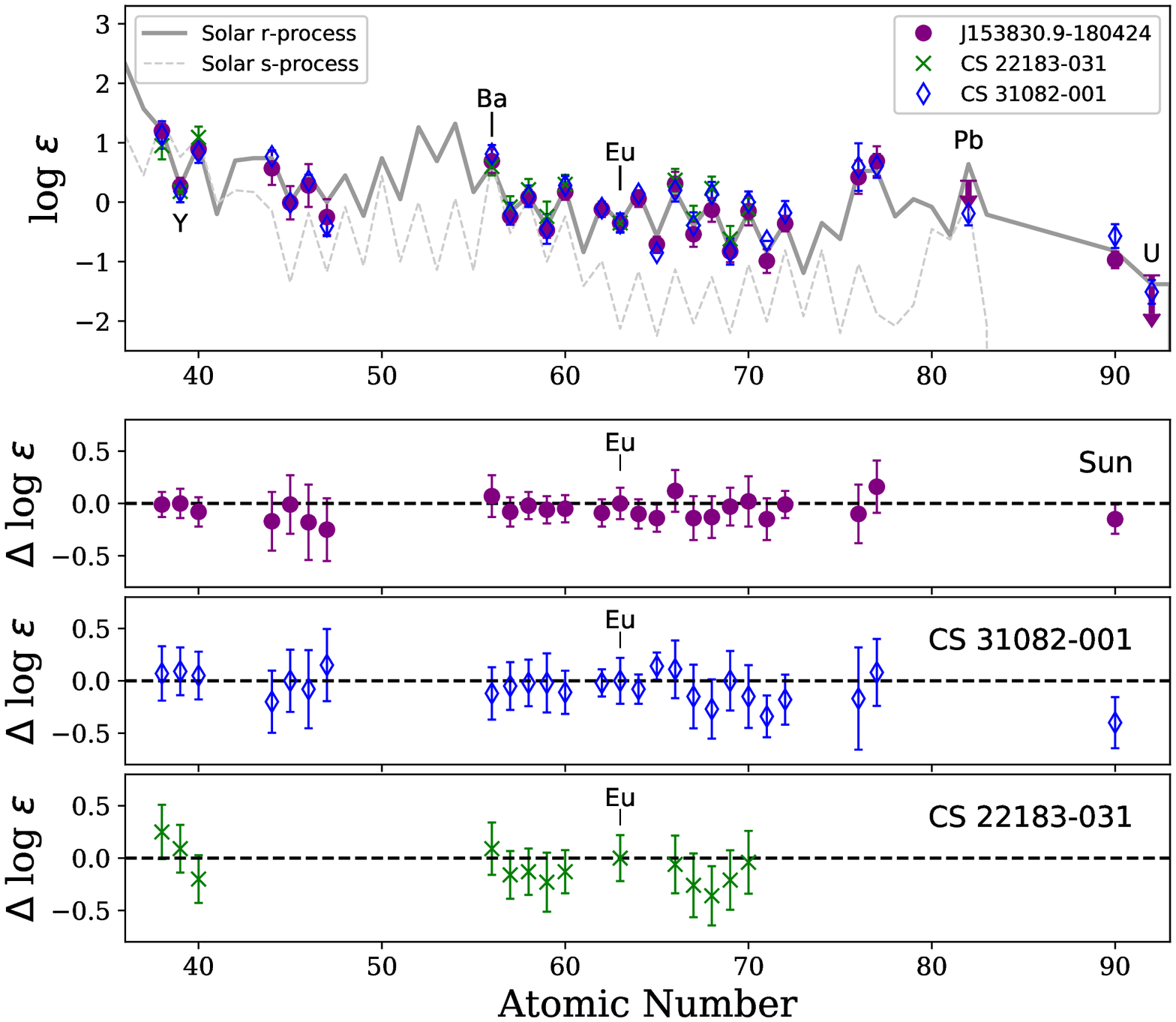}
\vspace{0.5in}
\caption{{\it Top panel: } Abundances of neutron-capture
  elements in J1538-1804 with the total errors (from Table
  \ref{table:Abunds}). Also shown are the $r$- and $s$-process patterns in
  the Sun (gray line, from \citealt{Arlandini1999}), and the
  abundances of an extremely metal-poor $r$-II (CS 31082-001; from
  \citealt{Hill2002}, \citealt{Sneden2009}, and
  \citealt{SiqueiraMello2013}) and $r$-I (CS 22183-031, from
  \citealt{Roederer2014b}) stars.  The solar $r$-process pattern and the
  metal-poor star abundances are shifted to the Eu abundance in
  J1538-1804; the solar $s$-process pattern is shifted to match the Ba
  abundance.  Upper limits are shown for Pb and U.  {\it Bottom
    panels: } Abundance offsets between J1538-1804 and the other
  stars, where $\Delta \log \epsilon(X) = \log \epsilon_{J1538}(X) -
  \log \epsilon_{star}(X).$  The offsets from the sun (second panel)
  are relative to the solar $r$-process
  residuals.\label{fig:rProcPatterns}}
\end{center}
\end{figure*}

Abundances or upper limits are derived for 27 neutron-capture
elements; sample syntheses are shown in Figure
\ref{fig:Spectra}. Table \ref{table:LineAbunds} shows the line-by-line
$r$-process abundances, while Table \ref{table:Abunds} shows the final
mean abundances.  When a spectral line was sufficiently weak and
unblended in the Kurucz solar
spectrum,\footnote{\url{http://kurucz.harvard.edu/sun.html}} the solar
abundance for that line was derived using the same atomic data (see
Table \ref{table:LineAbunds}; otherwise, the \citealt{Asplund2009} solar
values are used).  All [X/Fe] ratios are relative to the
[\ion{Fe}{1}/H] ratios in Table \ref{table:Targets} (note that the
adoption of NLTE corrections during the atmospheric parameter
determination has ensured that the \ion{Fe}{1} and \ion{Fe}{2} ratios
are equal).  The random errors are based on the quality of the
data and the fitted synthetic spectra.  A minimum random error of 0.05
dex was adopted, but weak or blended lines in low-S/N regions could
have random errors as large as $0.1-0.2$ dex.  The systematic errors
are due to uncertainties in the atmospheric parameters.  Table
\ref{table:Abunds} also provides the abundance offsets that
occur when LTE parameters are adopted.  Figure \ref{fig:rProcPatterns}
shows the $r$-process pattern in  J1538-1804, along with reference
patterns in the Sun, a very metal-poor, actinide boost $r$-II star
\citep{Hill2002}, and an extremely metal-poor $r$-I star
\citep{Roederer2014a}.  The pattern in J1538-1804 is well-fit by the
Solar $r$-process pattern, and does not agree with the Solar $s$-process
pattern.  The various groups of elements are discussed further below.

\startlongtable
\begin{deluxetable}{@{}lDcDcD}
\tabletypesize{\scriptsize}
\tablecolumns{9}
\tablewidth{0pt}
\tablecaption{Abundances per line.\label{table:LineAbunds}}
\hspace*{-4in}
\tablehead{
            & \multicolumn{2}{c}{Wavelength}  & EP     & \multicolumn{2}{c}{$\log gf$} & Solar &  \multicolumn{2}{c}{J1538-1804} \\
            & \multicolumn{2}{c}{(\AA)}       & (eV)   & \multicolumn{2}{c}{ }         & $\log \epsilon$ &  \multicolumn{2}{c}{$\log \epsilon$ }
}
\decimals
\startdata
\ion{Sr}{2} & 4161.792 & 2.938 & -0.50 & A09  & 1.20\pm0.05 \\
\ion{Y}{2}  & 3747.556 & 0.104 & -0.91 & A09  & 0.22\pm0.10 \\
\ion{Y}{2}  & 4398.013 & 0.129 & -1.00 & A09  & 0.22\pm0.10 \\
\ion{Y}{2}  & 4682.324 & 0.408 & -1.51 & A09  & 0.24\pm0.05 \\
\ion{Y}{2}  & 4883.680 & 1.083 &  0.07 & A09  & 0.42\pm0.10 \\
\ion{Zr}{2} & 3998.954 & 0.558 & -0.39 & A09  & 0.65\pm0.05 \\
\ion{Zr}{2} & 4050.316 & 0.713 & -1.00 & A09  & 1.02\pm0.05 \\
\ion{Zr}{2} & 4317.299 & 0.713 & -1.38 & A09  & 1.01\pm0.05 \\
\ion{Ru}{1} & 3436.736 & 0.148 & -0.02 & 1.67 & 0.51\pm0.10 \\
\ion{Ru}{1} & 3498.942 & 0.000 &  0.33 & 1.60 & 0.46\pm0.10 \\
\ion{Ru}{1} & 3798.898 & 0.148 & -0.09 & 1.75 & 0.56\pm0.10 \\
\ion{Ru}{1} & 3799.349 & 0.000 & -0.07 & 1.75 & 0.51\pm0.10 \\
\ion{Rh}{1} & 3396.819 & 0.000 &  0.05 & 0.91 & 0.07\pm0.10 \\
\ion{Rh}{1} & 3692.358 & 0.000 &  0.17 & 0.91 & -0.08\pm0.10 \\
\ion{Pd}{1} & 3404.579 & 0.813 &  0.32 & 1.37 & 0.08\pm0.10 \\
\ion{Ag}{1} & 3382.889 & 0.000 & -0.38 & 0.64 & 0.55\pm0.10 \\
\ion{Ba}{2} & 5853.675\tablenotemark{a} & 0.604 & -1.00 & A09 & 0.69\pm0.10 \\
\ion{Ba}{2} & 6141.713\tablenotemark{a} & 0.704 & -0.08 & A09 & 0.69\pm0.10 \\
\ion{La}{2} & 3988.515\tablenotemark{a} & 0.403 &  0.21 & A09 & -0.34\pm0.05 \\
\ion{La}{2} & 4086.709\tablenotemark{a} & 0.000 & -0.07 & A09 & -0.31\pm0.05 \\
\ion{La}{2} & 4123.230\tablenotemark{a} & 0.321 &  0.13 & A09 & -0.31\pm0.05 \\
\ion{La}{2} & 4333.750\tablenotemark{a} & 0.173 & -0.06 & A09 & -0.34\pm0.05 \\
\ion{La}{2} & 5303.520\tablenotemark{a} & 0.321 & -1.35 & A09 & -0.14\pm0.05 \\
\ion{La}{2} & 6390.460\tablenotemark{a} & 0.321 & -1.41 & A09 & -0.04\pm0.05 \\
\ion{Ce}{2} & 3940.330 & 0.318 & -0.27 & A09  & -0.14\pm0.05 \\
\ion{Ce}{2} & 3999.237 & 0.295 &  0.06 & A09  & -0.03\pm0.05 \\
\ion{Ce}{2} & 4014.897 & 0.529 & -0.20 & A09  & 0.09\pm0.05 \\
\ion{Ce}{2} & 4072.918 & 0.327 & -0.64 & A09  & 0.19\pm0.05 \\
\ion{Ce}{2} & 4073.474 & 0.477 &  0.21 & A09  & 0.07\pm0.05 \\
\ion{Ce}{2} & 4083.230 & 0.700 &  0.27 & A09  & 0.16\pm0.05 \\
\ion{Ce}{2} & 4120.840 & 0.320 & -0.37 & A09  & 0.19\pm0.05 \\
\ion{Ce}{2} & 4137.645 & 0.516 &  0.40 & A09  & -0.06\pm0.05 \\
\ion{Ce}{2} & 4138.096 & 0.924 & -0.08 & A09  & 0.17\pm0.05 \\
\ion{Ce}{2} & 4165.599 & 0.909 &  0.52 & A09  & 0.09\pm0.05 \\
\ion{Ce}{2} & 4222.597 & 0.122 & -0.15 & A09  & 0.09\pm0.05 \\
\ion{Ce}{2} & 4418.790 & 0.863 &  0.27 & A09  & 0.06\pm0.05 \\
\ion{Ce}{2} & 4449.330 & 0.608 &  0.04 & A09  & 0.04\pm0.05 \\
\ion{Ce}{2} & 4486.910 & 0.295 & -0.18 & A09  & 0.01\pm0.05 \\
\ion{Ce}{2} & 4562.370 & 0.477 &  0.21 & A09  & 0.09\pm0.05 \\
\ion{Ce}{2} & 4628.160 & 0.516 &  0.14 & A09  & 0.19\pm0.10 \\
\ion{Ce}{2} & 5274.230 & 1.044 &  0.13 & A09  & 0.19\pm0.05 \\
\ion{Pr}{2} & 3964.812\tablenotemark{a} & 0.055 &  0.07 & A09  & -0.52\pm0.05 \\
\ion{Pr}{2} & 4179.393\tablenotemark{a} & 0.204 &  0.46 & A09  & -0.40\pm0.05 \\
\ion{Pr}{2} & 4222.934\tablenotemark{a} & 0.055 &  0.23 & A09  & -0.48\pm0.05 \\
\ion{Pr}{2} & 4408.819\tablenotemark{a} & 0.000 &  0.05 & A09  & -0.48\pm0.05 \\
\ion{Pr}{2} & 5259.740\tablenotemark{a} & 0.633 &  0.12 & A09  & -0.47\pm0.10 \\
\ion{Nd}{2} & 3862.566 & 0.182 & -0.76 & A09  & 0.18\pm0.10 \\
\ion{Nd}{2} & 3863.408 & 0.000 & -0.01 & A09  & 0.03\pm0.10 \\
\ion{Nd}{2} & 3900.215 & 0.471 &  0.10 & A09  & 0.05\pm0.10 \\
\ion{Nd}{2} & 4021.728 & 0.182 & -0.31 & A09  & 0.18\pm0.05 \\
\ion{Nd}{2} & 4051.139 & 0.380 & -0.30 & A09  & 0.18\pm0.05 \\
\ion{Nd}{2} & 4061.080 & 0.471 &  0.55 & A09  & 0.23\pm0.05 \\
\ion{Nd}{2} & 4069.270 & 0.064 & -0.57 & A09  & 0.20\pm0.05 \\
\ion{Nd}{2} & 4177.320 & 0.064 & -0.10 & A09  & 0.12\pm0.05 \\
\ion{Nd}{2} & 4178.635 & 0.182 & -1.03 & A09  & 0.17\pm0.05 \\
\ion{Nd}{2} & 4179.580 & 0.182 & -0.64 & A09  & 0.13\pm0.10 \\
\ion{Nd}{2} & 4232.380 & 0.064 & -0.47 & A09  & 0.18\pm0.05 \\
\ion{Nd}{2} & 4446.390 & 0.204 & -0.35 & A09  & 0.15\pm0.05 \\
\ion{Nd}{2} & 4462.990 & 0.559 &  0.04 & A09  & 0.23\pm0.05 \\
\ion{Nd}{2} & 4959.120 & 0.064 & -0.80 & A09  & 0.25\pm0.05 \\
\ion{Nd}{2} & 4989.950 & 0.630 & -0.50 & A09  & 0.28\pm0.05 \\
\ion{Nd}{2} & 5092.790 & 0.380 & -0.61 & A09  & 0.19\pm0.05 \\
\ion{Nd}{2} & 5130.590 & 1.303 &  0.45 & A09  & 0.09\pm0.05 \\
\ion{Nd}{2} & 5212.350 & 0.204 & -0.96 & A09  & 0.23\pm0.10 \\
\ion{Nd}{2} & 5249.590 & 0.975 &  0.20 & A09  & 0.13\pm0.10 \\
\ion{Nd}{2} & 5319.820 & 0.550 & -0.14 & A09  & 0.23\pm0.05 \\
\ion{Sm}{2} & 3896.970 & 0.040 & -0.67 & A09  & -0.25\pm0.05 \\
\ion{Sm}{2} & 4188.128 & 0.543 & -0.44 & A09  & -0.05\pm0.05 \\
\ion{Sm}{2} & 4318.926 & 0.277 & -0.25 & A09  & -0.13\pm0.05 \\
\ion{Sm}{2} & 4421.126 & 0.378 & -0.49 & A09  & -0.03\pm0.05 \\
\ion{Sm}{2} & 4424.337 & 0.484 &  0.14 & A09  & -0.13\pm0.05 \\
\ion{Eu}{2} & 3724.931\tablenotemark{a} & 0.000 & -0.09 & 0.42 & -0.47\pm0.05 \\
\ion{Eu}{2} & 3907.107\tablenotemark{a} & 0.207 &  0.17 & 0.42 & -0.52\pm0.05 \\
\ion{Eu}{2} & 4129.725\tablenotemark{a} & 0.000 &  0.22 & 0.40 & -0.42\pm0.05 \\
\ion{Eu}{2} & 4435.578\tablenotemark{a} & 0.207 & -0.11 & 0.40 & -0.37\pm0.05 \\
\ion{Eu}{2} & 6645.064\tablenotemark{a} & 1.379 &  0.12 & 0.52 & -0.27\pm0.05 \\
\ion{Gd}{2} & 3549.359 & 0.240 &  0.29 & 0.97 & -0.32\pm0.10 \\
\ion{Gd}{2} & 3697.733 & 0.032 & -0.34 & 0.87 & -0.02\pm0.10 \\
\ion{Gd}{2} & 3768.396 & 0.078 &  0.21 & A09  & -0.03\pm0.10 \\
\ion{Gd}{2} & 3796.384 & 0.032 &  0.02 & A09  & 0.03\pm0.10 \\
\ion{Gd}{2} & 3844.578 & 0.144 & -0.46 & A09  & 0.23\pm0.10 \\
\ion{Gd}{2} & 4191.075 & 0.427 & -0.48 & A09  & 0.13\pm0.05 \\
\ion{Gd}{2} & 4215.022 & 0.427 & -0.44 & A09  & 0.13\pm0.05 \\
\ion{Gd}{2} & 4251.731 & 0.382 & -0.22 & A09  & 0.05\pm0.05 \\
\ion{Tb}{2} & 3702.850\tablenotemark{a} & 0.126 &  0.44 & A09  & -0.74\pm0.05 \\
\ion{Tb}{2} & 3747.380\tablenotemark{a} & 0.401 &  0.04 & A09  & -0.64\pm0.10 \\
\ion{Tb}{2} & 3848.730\tablenotemark{a} & 0.000 &  0.28 & A09  & -0.67\pm0.05 \\
\ion{Tb}{2} & 3874.168\tablenotemark{a} & 0.000 &  0.27 & A09  & -0.77\pm0.05 \\
\ion{Tb}{2} & 4002.566\tablenotemark{a} & 0.641 &  0.10 & A09  & -0.74\pm0.05 \\
\ion{Dy}{2} & 3757.368 & 0.100 & -0.17 & A09  & 0.01\pm0.10 \\
\ion{Dy}{2} & 3944.680 & 0.000 &  0.11 & A09  & 0.21\pm0.10 \\
\ion{Dy}{2} & 3996.689 & 0.590 & -0.26 & A09  & 0.26\pm0.05 \\
\ion{Dy}{2} & 4050.565 & 0.590 & -0.47 & A09  & 0.36\pm0.05 \\
\ion{Dy}{2} & 4073.120 & 0.540 & -0.32 & A09  & 0.38\pm0.05 \\
\ion{Dy}{2} & 4077.966 & 0.100 & -0.04 & A09  & 0.51\pm0.05 \\
\ion{Dy}{2} & 4103.306 & 0.100 & -0.38 & A09  & 0.46\pm0.05 \\
\ion{Ho}{2} & 3796.730\tablenotemark{a} & 0.000 &  0.16 & A09  & -0.44\pm0.05 \\
\ion{Ho}{2} & 3810.738\tablenotemark{a} & 0.000 &  0.19 & A09  & -0.56\pm0.10 \\
\ion{Ho}{2} & 3890.738\tablenotemark{a} & 0.079 &  0.46 & A09  & -0.61\pm0.10 \\
\ion{Er}{2} & 3692.649 & 0.055 &  0.28 & A09  & -0.12\pm0.10 \\
\ion{Er}{2} & 3729.524 & 0.000 & -0.59 & A09  & -0.02\pm0.10 \\
\ion{Er}{2} & 3786.836 & 0.000 & -0.52 & A09  & -0.07\pm0.10 \\
\ion{Er}{2} & 3830.481 & 0.000 & -0.22 & A09  & -0.17\pm0.10 \\
\ion{Er}{2} & 3896.233 & 0.055 & -0.12 & A09  & -0.07\pm0.10 \\
\ion{Er}{2} & 3906.311 & 0.000 &  0.12 & A09  & -0.17\pm0.10 \\
\ion{Er}{2} & 3938.626 & 0.000 & -0.52 & A09  & -0.27\pm0.10 \\
\ion{Tm}{2} & 3700.255 & 0.029 & -0.38 & A09  & -0.74\pm0.10 \\
\ion{Tm}{2} & 3701.362 & 0.000 & -0.54 & A09  & -0.89\pm0.10 \\
\ion{Tm}{2} & 3795.759 & 0.029 & -0.23 & A09  & -1.04\pm0.10 \\
\ion{Tm}{2} & 3848.019 & 0.000 & -0.14 & A09  & -0.67\pm0.10 \\
\ion{Tm}{2} & 3996.510 & 0.000 & -1.20 & A09  & -0.79\pm0.10 \\
\ion{Yb}{2} & 3694.190\tablenotemark{a} & 0.000 & -0.30 & 0.54 & 0.08\pm0.10 \\
\ion{Lu}{2} & 3507.380\tablenotemark{a} & 0.000 & -1.16 & 0.10 & -0.99\pm0.10 \\
\ion{Hf}{2} & 3719.276 & 0.608 & -0.81 & A09  & -0.34\pm0.10 \\
\ion{Hf}{2} & 3918.090 & 0.452 & -1.14 & A09  & -0.34\pm0.10 \\
\ion{Hf}{2} & 4093.150 & 0.452 & -1.15 & A09  & -0.39\pm0.05 \\
\ion{Os}{1} & 4135.775 & 0.515 & -1.26 & A09  & 0.56\pm0.05 \\
\ion{Os}{1} & 4260.848 & 0.000 & -1.44 & A09  & 0.31\pm0.05 \\
\ion{Os}{1} & 4420.520 & 0.000 & -1.53 & A09  & 0.38\pm0.05 \\
\ion{Ir}{1} & 3800.120 & 0.000 & -1.43 & A09  & 0.69\pm0.10 \\
\ion{Pb}{1} & 4057.807\tablenotemark{a} & 1.319 & -0.17 & A09  & <0.36 \\
\ion{Th}{2} & 4019.129 & 0.000 & -0.23 & A09  & -1.07\pm0.05 \\
\ion{Th}{2} & 4086.521 & 0.000 & -0.93 & A09  & -0.87\pm0.05 \\
\ion{U}{2}  & 3859.571 & 0.036 & -0.10 & A09  & <-1.23 \\
\enddata
\tablenotetext{a}{Line has hyperfine structure and/or isotopic splitting.}
\end{deluxetable}

\begin{deluxetable*}{@{}lcDccDcDcD}
\tabletypesize{\scriptsize}
\tablewidth{0pt}
\tablecaption{Mean Neutron-Capture Abundances and Errors.\label{table:Abunds}}
\hspace*{-4in}
\tablehead{
Element\tablenotemark{a} & $N$ & \multicolumn{2}{c}{\phantom{---}$\log \epsilon$\tablenotemark{b}} & $\sigma_{\rm{rand}}$ & $\sigma_{\rm{Total}}$\tablenotemark{c} &  \multicolumn{2}{c}{\phantom{---}$\sigma_{\rm{LTE}}$\tablenotemark{d}} & \phantom{sp} & \multicolumn{2}{c}{[X/\ion{Fe}{1}]} & $\sigma_{\rm{Total}}$\tablenotemark{c} & \multicolumn{2}{c}{\phantom{---}$\sigma_{\rm{LTE}}$\tablenotemark{d}}
}
\decimals
\startdata
\ion{Sr}{2} & 1  &  1.20 & 0.10 & 0.12 & -0.08 & & 0.44 & 0.20 & 0.07 \\
\ion{Y}{2}  & 4  &  0.27 & 0.05 & 0.14 & -0.11 & & 0.17 & 0.13 & 0.04 \\
\ion{Zr}{2} & 3  &  0.89 & 0.07 & 0.14 & -0.09 & & 0.47 & 0.13 & 0.06 \\
\ion{Ru}{1} & 5  &  0.57 & 0.05 & 0.28 &  0.10 & & 0.89 & 0.07 & 0.09 \\
\ion{Rh}{1} & 2  & -0.01 & 0.06 & 0.28 &  0.10 & & 1.15 & 0.08 & 0.09 \\
\ion{Pd}{1} & 1  &  0.28 & 0.10 & 0.36 &  0.07 & & 0.78 & 0.15 & 0.06 \\
\ion{Ag}{1} & 1  & -0.25 & 0.10 & 0.30 &  0.09 & & 0.88 & 0.11 & 0.08 \\
\ion{Ba}{2} & 2  &  0.69 & 0.07 & 0.20 & -0.16 & & 0.62 & 0.08 & -0.01 \\
\ion{La}{2} & 6  & -0.24 & 0.05 & 0.14 & -0.11 & & 0.77 & 0.12 & 0.04 \\
\ion{Ce}{2} & 17 &  0.08 & 0.05 & 0.13 & -0.07 & & 0.61 & 0.13 & 0.08 \\
\ion{Pr}{2} & 5  & -0.47 & 0.05 & 0.13 & -0.07 & & 0.92 & 0.13 & 0.08 \\
\ion{Nd}{2} & 20 &  0.17 & 0.05 & 0.13 & -0.08 & & 0.86 & 0.12 & 0.07 \\
\ion{Sm}{2} & 5  & -0.12 & 0.05 & 0.13 & -0.08 & & 1.03 & 0.12 & 0.07 \\
\ion{Eu}{2} & 5  & -0.32 & 0.05 & 0.15 & -0.16 & & 1.27 & 0.11 & -0.01 \\
\ion{Gd}{2} & 8  &  0.06 & 0.05 & 0.14 & -0.10 & & 1.10 & 0.12 & 0.05 \\
\ion{Tb}{2} & 5  & -0.71 & 0.05 & 0.13 & -0.09 & & 1.10 & 0.12 & 0.06 \\
\ion{Dy}{2} & 7  &  0.31 & 0.05 & 0.20 & -0.15 & & 1.32 & 0.07 & 0.0 \\
\ion{Ho}{2} & 3  & -0.54 & 0.05 & 0.21 & -0.15 & & 1.09 & 0.07 & 0.0 \\
\ion{Er}{2} & 7  & -0.13 & 0.05 & 0.20 & -0.14 & & 1.06 & 0.07 & 0.01 \\
\ion{Tm}{2} & 5  & -0.83 & 0.05 & 0.18 & -0.08 & & 1.18 & 0.08 & 0.07 \\
\ion{Yb}{2} & 1  &  0.25 & 0.10 & 0.24 & -0.24 & & 1.52 & 0.11 & -0.09 \\
\ion{Lu}{2} & 1  & -0.99 & 0.10 & 0.20 & -0.06 & & 0.92 & 0.12 & 0.09 \\
\ion{Hf}{2} & 3  & -0.36 & 0.05 & 0.13 & -0.06 & & 0.90 & 0.13 & 0.09 \\
\ion{Os}{1} & 3  &  0.42 & 0.05 & 0.28 &  0.09 & & 1.13 & 0.07 & 0.08 \\
\ion{Ir}{1} & 1  &  0.44 & 0.10 & 0.25 &  0.02 & & 1.17 & 0.10 & 0.01 \\
\ion{Pb}{1} & 1  & <0.36 &      &      & \multicolumn{2}{c}{ }      & & <0.72 &     & \multicolumn{2}{c}{ }\\
\ion{Th}{2} & 2  & -0.97 & 0.07 & 0.14 & -0.05 & & 1.12 & 0.14 & 0.10 \\
\ion{U}{2}  & 1  & <-1.23 &     &      & \multicolumn{2}{c}{ }      & & <1.42 &     & \multicolumn{2}{c}{ }\\
\enddata
\tablenotetext{a}{In order of atomic number.}
\tablenotetext{b}{A mean [X/H] abundance was calculated
  with a straight mean of the differential [X/H] ratios (using
  the solar values in Table \ref{table:LineAbunds}); the mean [X/H]
  was then converted to $\log \epsilon$ with the \citet{Asplund2009}
  solar abundance.}
\tablenotetext{c}{The total error refers to the combination of random
  and systematic errors (where the latter are due to uncertainties in
  the atmospheric parameters), calculated according to Equations A1,
  A4, and A5 in \citet{McWilliam2013}.  Errors in $\log \epsilon$ and
  [X/Fe] are listed separately.}
\tablenotetext{d}{The LTE error refers to the offset that results when
LTE atmospheric parameters are used (see Paper I).  Errors in $\log
\epsilon$ and [X/Fe] are listed separately.  These offsets are not included in $\sigma_{\rm{Total}}$.}
\end{deluxetable*}

\subsection{Barium and Europium}\label{subsec:BaEu}
Barium and europium are the elements used for classification of $r$-I
and $r$-II stars.  The Ba lines are quite strong in this fairly
metal-rich star; only the 5853 and 6141 \AA \hspace{0.025in} lines
were used.  Multiple \ion{Eu}{2} lines were detected, including the
weak line at 6645 \AA \hspace{0.025in}.  The derived subsolar value of
[Ba/Eu] ($-0.65\pm0.08$) suggests that the star has minimal
contamination from the s-process despite its moderate [Fe/H], while
its high [Eu/Fe] ($+1.27\pm 0.05$) makes it an $r$-II star.

J1538-1804 is on the metal-rich end of known $r$-II stars.  There are
only four $r$-II stars in JINAbase that have $[\rm{Fe/H}]\ga -2.1$,
and only one has a detailed $r$-process pattern that covers from Sr to
U and provides an age determination.  Two of these four stars are
associated with dwarf galaxies (Reticulum~II and Ursa Minor;
\citealt{Ji2016,Aoki2007}), while the other two are associated with
the Milky Way bulge \citep{Howes2016}. J1538-1804 is therefore the
most metal-rich $r$-II star known in the Milky Way halo.

\subsection{Lighter $r$-Process Elements}\label{subsec:LightR}
The lighter $r$-process elements Sr, Y, and Zr were derived
with 1, 4, and 6 lines, respectively (see Figure \ref{fig:Spectra} for
the fit to the Sr line).  These elements are commonly derived in many
abundance analyses.   Abundances of Ru and Rh were determined from 5
and 2 lines, while Pd and Ag were derived from single lines in the
noisier blue end of the spectrum.  Sr, Y, and Zr are in agreement
with the Solar pattern and the other $r$-I and $r$-II stars.   Ru and
Ag are slightly low in J1538-1804 compared to the Sun; CS 31082-001
also has similarly low Ag relative to the Sun.

These lighter $r$-process elements are thought to form in the main
$r$-process and in a ``weak'' or ``limited'' $r$-process (Frebel {\it in
  prep.}) that could occur in traditional core-collapse supernovae
(e.g., \citealt{ArconesThielemann2013}). The lighter $r$-process
pattern may therefore vary between stars depending on age,
environment, etc. Both \citet{SiqueiraMello2014} and
\citet{Placco2017} have noted first-peak variations in $r$-I and
$r$-II stars.  J1538-1804's abundance pattern is robust with respect
to the Sun, suggesting that the relative contribution from the
``limited'' and main $r$-processes are similar.

\subsection{The Lanthanides and Hafnium}\label{subsec:Lanthanides}
There are many detectable, relatively unblended lines from rare earth
elements (La through Lu) and Hf.  There is only a single, strong line
available for Yb, which requires hyperfine and isotopic structure.
The pattern of these elements agrees well with the Sun and with the
other $r$-I and $r$-II stars.  This robust pattern amongst the
rare earth elements is a feature that is seen in all
$r$-process-enhanced stars (e.g., \citealt{Sneden2008}), and is
therefore an essential requirement for models of $r$-process
nucleosynthesis.

\subsection{Third-Peak Elements}\label{subsec:OsIr}
Os and Ir are the only third-peak elements available in this
spectrum.  Os has three lines, though Ir only has one.  Despite the
paucity of lines, both elements agree very well with the other
patterns in Figure \ref{fig:rProcPatterns} (and with other $r$-II stars;
e.g., \citealt{Placco2017}).

The relative strength of the second and third $r$-process peaks is
also important for constraining the physical conditions of the
$r$-process (see Section \ref{subsec:Pattern}). In J1538-1804, in the
other $r$-I and $r$-II stars, and in the Sun, the pattern between the
second and third peaks appears to be consistent across $\sim 4$ dex in
metallicity.

\subsection{Lead}\label{subsec:Pb}
Only an upper limit of $\log \epsilon(\rm{Pb}) = 0.36$ can be derived
from the line at 4057.8 \AA.  Pb is a significant product of the
$s$-process.  This upper limit therefore supports the assertion from the
[Ba/Eu] that any contamination from the $s$-process in this star is
minimal.  This agrees with \citet{Roederer2010}, who find minimal
$s$-process contributions to MW stars with metallicities as high as
$[\rm{Fe/H}] \sim -1.4$.

\subsection{The Actinides}\label{subsec:ThU}
Th and U are highly desirable elements for $r$-process studies for two
reasons: 1) A handful of $r$-II stars exhibit an enhancement in the
actinides (e.g., {\color{red} \citealt{Schatz2002}}) and 2) U and Th are
radioactive, and relative abundance ratios with respect to stable
elements like Eu are therefore useful for age-dating (see Section
\ref{subsec:Age}).

There are two clean \ion{Th}{2} lines in this star, enabling a
high-precision Th measurement (see Figure \ref{fig:Spectra}).  The
strongest \ion{Th}{2} line at 4019 \AA \hspace{0.025in} is often
blended, but this problem is reduced at low metallicities.  The
\ion{U}{2} line at 3859 \AA \hspace{0.025in} is severely blended with
an \ion{Fe}{1} in J1538-1804, providing only an upper limit.  Based on
its Th and U abundances, J153830.9-180424 does not appear to be an
actinide boost star.


\section{Discussion}\label{sec:Conclusion}

\subsection{The $r$-Process Pattern}\label{subsec:Pattern}
Despite its high metallicity relative to other $r$-II stars,
J1538-1804 is a typical $r$-II star.  Its $r$-process abundance
pattern matches that of the Sun and the $r$-II star CS~31082-001
(except that J1538-1804 does not exhibit an actinide boost like
CS~31082-001). Its Pb abundance and [Ba/Eu] ratio suggest that the
$s$-process has not contributed to its abundance pattern.  It does not
exhibit the light $r$-process variations that have been observed in
other $r$-I and $r$-II stars (e.g., \citealt{Honda2006}), and
therefore does not require excessive contributions from the
``limited'' (or ``weak'') $r$-process (see Frebel et al., {\it in
  prep.}).

Patterns such as these are essential for identifying the site(s) of
the $r$-process.  The abundance patterns from models are highly
sensitive to uncertainties in nuclear masses, $\beta$-decay rates,
fission cycling, neutrino properties, etc. (e.g.,
\citealt{Surman2017}), as well as the physical conditions of the
environment, such as temperature, the electron fraction, and the
density \citep{Hoffman1997}.  The pattern in J1538-1804 sets strong
requirements for $r$-process models.

\subsection{Age}\label{subsec:Age}
The Th abundance and the upper limit on U indicate that
J153830.9-180424 is an old star. The relative abundances between
Th and all the second- and third- peak elements in Table
\ref{table:Abunds} give a mean age of $11.2\pm3.9$ Gyr when the 
\citet{Schatz2002} initial production ratios (from waiting-point
calculations) are adopted, and $17.2\pm7.2$ Gyr when the
\citet{Hill2016} values (from a high-entropy wind model) are adopted
(see Table 7 in \citealt{Placco2017}).  The quoted uncertainties
represent the standard deviations from different chronometer
pairs. The upper limit in the U abundance yields a lower limit from
U/Eu of $5.4-5.7$ Gyr, depending on the production ratio.  The
dominant sources of uncertainty in these ages are the choices of
production ratios and chronometer pairs; while U/Th would be a better
ratio, the upper limit on U is not very constraining in this case.
These ratios strongly suggest that the $r$-process material in
J1538-1804 is indeed ancient.

\subsection{J1538-1804 and the Site of the $r$-Process}\label{subsec:BigPicture}
Given that J1538-1804 is a typical $r$-II star, what is gained from
these observations?  First, this adds another $r$-II star to the known
sample of $\sim~30$, an important step for statistically analyzing the
$r$-process patterns as a function of stellar properties (metallicity,
age, location in the Galaxy, etc.).  Secondly, this letter has
demonstrated that the main $r$-process pattern at $[\rm{Fe/H}]\sim-2$ is
very similar to the pattern at $[\rm{Fe/H}]\sim-4$ and
$[\rm{Fe/H}]\sim0$, a powerful result when combined with the old age
of the $r$-process material.  This either implicates a single site for
the $r$-process, or requires that $r$-process nucleosynthesis in different
sites leads to the same final abundance pattern.

The very metal-poor $r$-I and $r$-II stars {\it without} $s$-process
contamination also provide constraints on the birth environments of
these stars, such as star formation rates, timescales relative to the
onset of type Ia supernovae, etc.  The recent discovery of
$r$-process-enhanced stars in ultra-faint dwarfs (e.g.,
\citealt{Ji2016}) has led to speculation that all
$r$-process-enhanced stars may originate in dwarf galaxies.
Indeed, J1538-1804's retrograde orbit in the MW halo does hint at a
possible extragalactic origin.  The chemical enrichment of lower mass
systems proceeds more slowly than in more massive galaxies; in
particular, dwarf galaxies cannot form as many metal-rich stars, and
the onset of the $s$-process occurs at a lower [Fe/H] than in the Milky
Way (e.g., \citealt{Tolstoy2009}). As more $r$-I and $r$-II stars are
discovered, particularly at higher metallicities, the general
properties of the $r$-process-enhanced stellar population will place
limits on the nature of the birth environments of these stars.

In the near future, the R-Process Alliance will provide data for many
more $r$-process-enhanced stars.  This will enable $r$-I and $r$-II
stars to be studied as stellar populations, and will provide $r$-process
patterns that can be used to tease out any variations as a function of
stellar properties such as metallicity, location, and more.

\acknowledgements
The authors wish to the thank the anonymous referee for helpful comments.
C.M.S., G.W., C.E.D., and E.M.F. acknowledge funding from the Kenilworth
Fund of the New York Community Trust.
V.M.P., E.M.H., T.C.B., and I.U.R. acknowledge partial support from grant
PHY 14-30152 (Physics Frontier Center/JINA-CEE), awarded by the US
National Science Foundation.
Australian access to the Magellan Telescopes was supported through the
National Collaborative Research Infrastructure Strategy of the
Australian Federal Government. 

\footnotesize{

}


\begin{thebibliography}{99}
\bibitem[Abbott et al.(2017)] {Abbott2017} Abbott, B.P., Abbott, R., Abbott, T.D., et al. 2017, {\it PhRvL}, 119, 1101
\bibitem[Abohalima et al.(2017)] {Abohalima2017} Abohalima, A. \&
  Frebel, F. 2017, arXiv:1711.04410
\bibitem[Amarsi et al. (2016)] {Amarsi2016} Amarsi, A.M., Lind, K.,
  Asplund, M., Barklem, P.S., \& Collet, R. 2016, \mnras, 463, 1518
\bibitem[Aoki et al.(2007)] {Aoki2007} Aoki, W., Honda, S., Sadakane, K., \& Arimoto, N. 2007, \pasj, 59, 15
\bibitem[Arcones \& Thielemann (2013)] {ArconesThielemann2013}
  Arcones, A., \& Thielemann, F.-K. 2013, {\it Journal of Physics G:
    Nuclear and Particle Physics}, 40, 3201
\bibitem[Arlandini et al.(1999)] {Arlandini1999} Arlandini, C.,
  K\"{a}ppeler, F., Wisshak, K., Gallino, R., Lugaro, M., Busso, M.,
  Straniero, O. 1999, \apj, 525, 886
\bibitem[Asplund et al.(2009)] {Asplund2009} Asplund, M., Grevesse,
N., Sauval, J.A., \& Scott, P. 2009, \araa, 47, 481
\bibitem[Barklem et al.(2005)] {Barklem2005} Barklem, P.S.,
  Christlieb, N., Beers, T.C., et al. 2005, \aap, 439, 129
\bibitem[Beniamini et al.(2016)] {Beniamini2016} Beniamini, P.,
  Hotokezaka, K., \& Piran, T. 2016, \apj, 832, 149
\bibitem[Bernstein et al.(2003)] {Bernstein2003} Bernstein, R.,
  Shectman, S. A., Gunnels, S. M., Mochnacki, S., \& Athey,
  A. E. 2003, in Society of Photo-Optical Instrumentation Engineers
  (SPIE) Conference Series, Vol. 4841, Society of Photo-Optical
  Instrumentation Engineers (SPIE) Conference Series, ed. M. Iye \&
  A. F. M. Moorwood, 1694
\bibitem[Brown et al.(2016)] {GAIAref} Brown, A.G.A., Vallenari, A.,
  Prusti, T., et al. 2016, \aap, 595, A2
\bibitem[Cayrel et al.(2001)] {Cayrel2001} Cayrel, R., Hill, V.,
  Beers, T.C., et al. 2001, \nat, 409, 691
\bibitem[Chornock et al.(2017)] {Chornock2017} Chornock, R., Berger,
  E., Kasen, D., et al. 2017, \apj, 848, L19
\bibitem[Christlieb et al.(2004)] {Christlieb2004} Christlieb, N.,
  Beers, T.C., Barklem, P.S., et al. 2004, \aap, 428, 1027
\bibitem[C\^{o}t\'{e} et al.(2017)] {Cote2017} C\^{o}t\'{e}, B.,
  Fryer, C.L., Belczynski, K., et al. 2017, arXiv:1710.05875
\bibitem[Cowan et al.(2005)] {Cowan2005} Cowan, J.J., Sneden, C., Beers, T.C., et al. 2005, \apj, 627, 238
\bibitem[Frebel et al.(2015)] {Frebel2015} Frebel, A., Chiti, A., Ji,
  A.P., Jacobson, H.R., \& Placco, V.M. 2015, \apjl, 810, L27
\bibitem[Hill et al.(2002)] {Hill2002} Hill, V., Plez, B., Cayrel, R.,
  et al. 2002, \aap, 387, 560
\bibitem[Hill et al.(2017)] {Hill2016} Hill, V., Christlieb, N., \&
  Beers, T. C., et al. 2017, \aap, 607, 91
\bibitem[Hoffman et al.(1997)] {Hoffman1997} Hoffman, R.D., Woosley,
  S.E., \& Qian, Y.-Z. 1997, \apj, 482, 957
\bibitem[Honda et al.(2006)] {Honda2006} Honda, S., Aoki, W.,
  Ishimaru, Y., Wanajo, S., \& Ryan, S.G. 2006, \apj, 643, 1180
\bibitem[Howes et al.(2016)] {Howes2016} Howes, L.M., Asplund, M., Keller, S.C., et al. 2016, \mnras, 460, 884
\bibitem[Ji et al.(2016)] {Ji2016} Ji, A.P., Frebel, A., Simon, J.D.,
  \& Chiti, A. 2016, \apj, 830, 93
\bibitem[Lattimer \& Schramm(1974)] {LattimerSchramm1974} Lattimer,
  J. \& Schramm, D. 1974, \apjl, 192, L145
\bibitem[Lippuner et al.(2017)] {Lippuner2017} Lippuner, J., Fern\'{a}ndez, R., Roberts, L.F., Foucart, F., Kasen, D., Metzger, B.D., \& Ott, C.D. 2017, \mnras, 472, 904 
\bibitem[Macias \& Ramirez-Ruiz(2016)] {MaciasRamirezRuiz2016} Macias,
  P. \& Ramirez-Ruiz, E. 2016, arXiv:1609.04826
\bibitem[Matijevi\u{c} et al.(2017)] {Matijevic2017} Matijevi\u{c},
  G., Chiappini, C., Grebel, E.K., et al. 2017, \aap, 603, 19
\bibitem[McWilliam et al.(2013)] {McWilliam2013} McWilliam, A.,
  Wallerstein, G., \& Mottini, M. 2013, \apj, 778, 149
\bibitem[Placco et al.(2014)] {Placco2014c} Placco, V.M., Frebel, A.,
  Beers, T.C., \& Stancliffe, R.J. 2017, \apj, 797, 21
\bibitem[Placco et al.(2017)] {Placco2017} Placco, V.M., Holmbeck,
  E.M., Frebel, A., et al. 2017, \apj, 844, 18
\bibitem[Roederer(2013)] {Roederer2013} Roederer, I.U. 2013, \aj, 145,
  26
\bibitem[Roederer et al.(2010)] {Roederer2010} Roederer, I.U., Cowan,
  J.J., Karakas, A.I., et al. 2010, \apj, 724, 975
\bibitem[Roederer et al.(2014a)] {Roederer2014a} Roederer, I.U.,
  Cowan, J.J., Preston, G.W., et al. 2014a, \mnras, 445, 2970
\bibitem[Roederer et al.(2014b)] {Roederer2014b} Roederer, I.U.,
  Preston, G.W., Thompson, I.B., et al. 2014b, \aj, 147, 136
\bibitem[Rosswog et al.(2014)] {Rosswog2014} Rosswog, S., Korobkin,
  O., Arcones, A., Thielemann, F.-K., Piran, T. 2014, \mnras, 439, 744
\bibitem[Schatz et al.(2002)] {Schatz2002} Schatz, H., Toenjes, R.,
  Pfeiffer, B., Beers, T.C., Cowan, J.J., Hill, V., \& Kratz,
  K.-L. 2002, \apj, 579, 626
\bibitem[Siqueira Mello et al.(2013)] {SiqueiraMello2013} Siqueira
  Mello, C., Spite, M., Barbuy, B. 2013, \aap, 550, 122
\bibitem[Siqueira Mello et al.(2014)] {SiqueiraMello2014} Siqueira
  Mello, C., Hill, V., Barbuy, B., et al. 2014, \aap, 565, 93
\bibitem[Sneden(1973)]{Sneden} Sneden, C. 1973, \apj, 184, 839
\bibitem[Sneden et al.(1994)]{Sneden1994} Sneden, C., Preston, G.W.,
  McWilliam, A., \& Searle, L. 1994, \apj, 431, 27
\bibitem[Sneden et al.(1996)] {Sneden1996} Sneden, C., McWilliam, A.,
  Preston, G.W., Cowan, J.J., Burris, D.L., \& Armosky, B.J. 1996,
  \apj, 467, 819
\bibitem[Sneden et al.(2008)] {Sneden2008} Sneden, C., Cowan, J.J., \&
  Gallino, R. 2008, \araa, 46, 241
\bibitem[Sneden et al.(2009)] {Sneden2009} Sneden, C., Lawler, J.E., Cowan, J.J., Ivans, I.I., \& Den Hartog, E.A. 2009, \apjs, 182, 80
\bibitem[Sobeck et al.(2011)] {Sobeck2011} Sobeck, J. S., Kraft,
  R. P., Sneden, C., et al. 2011, \aj, 141, 175
\bibitem[Steinmetz et al.(2006)] {RAVEref} Steinmetz, M., Zwitter, T.,
  Siebert, A., et al. 2006, \aj, 132, 1645
\bibitem[Surman et al.(2017)] {Surman2017} Surman, R., Mumpower, M.,
  \& McLaughlin, G. 2017, Proc. of the 14th International Symposium on
  Nuclei in the Cosmos, ed. Kubono, S., Kajino, T., Nishimura, S.,
  Isobe, T., Nagataki, S. Shima, T., \& Takeda, Y.
\bibitem[Tolstoy et al.(2009)] {Tolstoy2009} Tolstoy, E., Hill, V., \&
Tosi, M. 2009, \araa, 47, 371
\bibitem[Tsujimoto \& Nishimura(2015)] {TsujimotoNishimura2015}
  Tsujimoto, T. \& Nishimura, N. 2015, \apjl, 811, L10
\bibitem[Winteler et al.(2012)]{Winteler2012} Winteler, C.,
  K\"{a}ppeli, R., Perego, A., Arcones, A., Vasset, N., Nishimura, N.,
  Liebend\"{o}rfer, M., \& Thielemann, F.-K. 2012, \apjl, 750, L22
\bibitem[Xu et al.(2007)] {Xu2007} Xu, H.L., Svanberg, S., Quinet, P., Palmeri, P., \& Bi\'{e}mont, \'{E}. 2007, {\it JQSRT}, 104, 52
\end{thebibliography}
\end{document}